# Design strategies for empathetic AI robots for older adults

Isabel Pedersen, Ontario Tech University and Andrea Slane, Ontario Tech University

*Abstract*— **Emulating empathy in human-robot interaction is a key component for achieving satisfying social, trustworthy, and ethical robot interaction with older people. Following comments from older adult study participants, the article uses humanities methods to identify a gap in defining empathetic robot care activities. It provides a design focus to mitigate it. Current human-robot designs, to a certain extent, neglect to include empathy as a theorized design pathway. Using one digital humanities research collection on humanoid robots, it contributes an empathetic care vocabulary as a design pathway for a productive underlying foundation for designing Socially Assistive Robots (SARs) that aim to support older people's goals of aging-in-place. Using rhetorical theory, this paper defines the socio-cultural expectations for convincing empathetic relationships.**

## I INTRODUCTION

Can robots be empathetic? Can social robot design evolve enough for humans to see robots as effective confidantes, companions, significant others or close personal friends rather than automatons? This broad question is relevant to different demographics of people for different reasons. Posed for older people, these questions take on weight given that social robots are marketed to them with urgency to address social isolation and loneliness. Robots are thought of as a salve for the global aging crisis [1]. The *robots as companions* theme has already gained a foothold to support aging populations across the globe [2] [3] [4] [5] [6] [7] [8]. Oftentimes, however, the goal to save the future from the onslaught of aging is a discourse of urgency that binds social robots to a value system that is more about efficiency and shortfalls in human social support, than empathy.

The business of 'assuaging loneliness' has become commercial in many ways. AI chatbots are proliferating in the form of AI platonic or romantic friendship apps (e.g., Replika, character.ai, Candy.ai, Kupid AI, Nectar AI). It appears that the next wave of social robots will use both mainstream virtual assistants (e.g., Apple Siri, Amazon Alexa) and AI chatbots to enable dynamic conversation in multiple languages.

Yet the challenge is rooted in achieving empathy. If robot 'content' designers and developers can strive to create empathetic systems that use methods based on human empathy as a value system – not simply decontextualized conversations and vacuous or sycophantic responses – the likelihood of a social robot appearing to understand the feelings of an older person and hence more effectively counter loneliness also becomes more likely in synthetic emotional interactions. This article attends to calls from the field, that "developers must collaborate with social scientists, medical professionals, and end-users to address safety features and ethical concerns surrounding social robots" [2].

We concentrate on the most developed category of robots, *socially assistive robots (SARs)*, which are marketed to older people. They are defined for their assistive abilities, e.g., task-focused 'care' activities, security/safety assistance and their conversational 'social' functions [9]. SARs are "created to meet challenges of the global increase of older adults. SARs are autonomous embodied technologies, equipped with auditory and visual faculties, enabling them to interact with users while performing assistive roles" [9]. Of all categories of social robots, they are meant to care for people and be empathetic.

For this article, however, we also draw on the broader *cultural* categories of social robots, care robots, humanoid robots, or even toy robots in the methods because defining social values for empathy overlaps with SARs in discursive arrangements. We understand that these categories are defined differently in science and engineering fields, yet the design focus and ultimately the goal to define ideal traits can overlap discursively when striving for empathetic Human-Robot Interaction [HRI].

The article contributes:

(1) a **novel humanities-informed methodology** to help achieve empathetic-appearing robots,

(2) a novel and relevant **empathetic care vocabulary** that could be used by other designers/developers.

(3) **design concepts and design traits** for developers to help achieve socially assistive robots that behave with empathy

Section Two explains background work relevant to this article. Section Three provides a narrative literature review of relevant articles from various disciplines, placing deliberate attention on articles that have contributed to the field of empathetic robots both in computer science and outside of it. Section Four describes qualitative research methods used for the article. Section Five discusses the use of digital humanities tools to analyze the discursive treatment of robot care, both through cultural artifacts discussed across different media and those reported by robot companies. Section Six defines and contextualizes the concept of empathy as a social value system using rhetoric as a methodological approach. This section positions empathy's discursive framing and offers three ideal rhetorical design concepts for the design of empathetic robots.

## II Background Work

While robots might appear to behave socially, the extent to which a robot *should* serve as a surrogate for human social interaction and a remedy for loneliness is still highly debatable and controversial [10] [11]. This is an issue we point to throughout this article because it impacts ethical alignment. Another issue is the lack of humanities and social sciences research in both the design and evaluation of AI assistants and social robots. One report states the need to invest in social abilities directly: "given that the current landscape of AI evaluation focuses primarily on the technical components of AI systems, it is important to invest in the holistic sociotechnical evaluations of AI assistants, including human–AI interaction, multi-agent and societal level research, to support responsible decision-making and deployment in this domain" [12, p. i]. This article responds with one such 'holistic' approach to empathy as a value system.

Along the lines of sociotechnical and socioaffective alignment [13], we previously conducted a multiyear research study on how currently independently living older people think about the prospect of bringing social robots into their homes and it revealed inspired prescriptions for future robot designs [14] [1]. With research ethics board approval, the study included focus groups and a survey of older adults at a time when social distancing restrictions imposed to help prevent spread of the COVID-19 pandemic were in force. One question in the survey asked if their pandemic experience had affected their opinions about using humanoid robots, and if yes, how have they changed?

A common sentiment was that some respondents were more receptive to the idea after the pandemic, but with reservations. In sixty-one responses, the overwhelming agreement was that human contact was superior to humanoid robots. However, responses also indicated the need to do a better job of situating humanoid robots within a social ecosystem to move the needle farther toward providing a benefit.

To get there, empathetic-behaving robots, this paper focuses on empathy as a core human value and distills ethical design concepts and traits for social robots with AI capabilities to mitigate a gap in the research.

## III Literature Review

The goal for this narrative literature review is to provide an overview of how relevant fields are contributing key insights to the goal of creating an empathetic appearing robot for older people living at home. This section does not claim to provide a systematic review of the literature but rather aims to identify multidisciplinary opportunities to fill research gaps. As such, this review touches not only on previous computer science fields including HRI and affective computing, but also deliberately includes work from media studies, aging studies, philosophy and ethics, and communication studies to enable an interdisciplinary approach.

### *A Empathetic computing*

Empathetic computing for HRI or other kinds of HCI is an established computer science field [15] [16] [17]. Robots that *read* human emotions, *perform* human emotions, and *behave* ethically are in demand now.

Long-standing research on machine ethics and specifically, robot ethics, needs to inform empathetic AI robot design [18]. One foundational paper identifies the category, "Full Ethical Agents" that can make "explicit moral judgments and justify them" [18, p. 511]. This important category presupposes the capacity for AI agents embedded in robots to make decision autonomously. It pinpoints the risk, and one value for empathy, that "society as a whole needs to take responsibility for the human causalities of accidents caused by such machines" [18, p. 514].

If AI is to appear empathetic, it must be programmed to make choices or react to humans in a manner that is also ethical. Another team of researchers proposes methods for designing autonomous robots that would be categorized according to degrees of moral agency, including one that can "behave in a way that shows an understanding of responsibility to some other moral agent" [19, p. 588].

*B Empathetic computing standards*

The IEEE Standard for Ethical Considerations in Emulated Empathy in Autonomous and Intelligent Systems report defines “empathic autonomous and intelligent system(s)” as “affect-sensitive technologies employed to algorithmically infer, model, simulate, or stimulate understanding of emotions, feelings, moods, perspective, attention or intention.” [20, p. 21). The emphasis is on robot behaviour simulated through computing technology. Outcomes for affected stakeholders are considered, and the standard notes the problem of making assumptions about individuals. However, it also notes, “the meaning of the word empathy remains ill-defined.” [20, p. 43]

In the field of synthetic emotions engineering standards, empathy is assigned an important, foundational role: “deliberately constructed emotions are designed **to create empathy between humans and artifacts, which may be useful or even essential for human-A/IS** [Autonomous and Intelligent Systems] collaboration” [emphasis added] [21, p. 105]. In this manner, the other emotions (the appearance of love, anger, or fear etc.) will be used to establish empathy.

The IEEE recommendation that AI “should not cause harm by either amplifying or dampening human emotional experience,” [21, p. 6] acknowledges the strong potential for humans to have emotional attachments with AI and the realization that emotions can be augmented, stifled, or misdirected to the detriment of humans. However, a thorough treatment of empathy itself as a human value is not included in the standard.

Affective Computing involves the capacity for computer devices to interpret, process, and simulate human affective experiences or feelings, emotions, or moods [21, p. 180]. To address the ethical concerns, AI developers have called for better methods to guide ethical design of emotional exchange for affective computing: “Affect is a core aspect of intelligence. Drives and emotions such as anger, fear, and joy are often the foundations of actions throughout our lives. To ensure that intelligent technical systems will be used to help humanity to the greatest extent possible in all contexts, autonomous and intelligent systems that participate in or facilitate human society should not cause harm by either amplifying or dampening human emotional experience” [21, p. 6].

The IEEE Standard for Ethical Considerations in Emulated Empathy in Autonomous and Intelligent Systems states that “training the system with empathy involves developing an understanding of the subject’s affect, as well as the surrounding context, such as cultural and individual differences that can influence emotional expression” [20, p. 36]. At the same time, while “Emulated empathy may enhance usability, engagement, and accessibility, . . . it also raises concerns about manipulation, commodification of interior life, and detachment from reality.” [22]

Human rights are at stake when synthetic emotions are developed. As an IEEE report on affective computing notes, guidelines for appropriate use of autonomous and intelligent affective systems “should acknowledge fundamental human rights to highlight potential ethical benefits and risks that may emerge, if and when affective systems interact intimately with users” [21, p. 95]. Designing robots that promise empathy require designers to incorporate safeguards against ethical hazards into their designs.

*C Empathetic social robots for aging participants*

Artificial Intelligence (AI) techniques like Machine Learning have led to speech processing functions, which have enabled robots to have extensive conversational abilities for social interaction [23] with older people. In the substantial literature on robots for aged care, many describe development of social functionalities, whether they serve to encourage human participants to connect socially with each other or to encourage humans to engage socially with robots [2]. Empathy is deemed necessary for SARs, especially in medical and residential care settings, as a foundation for them to behave as trusted social supports [24] [25] [26]. Another team analyzes empathy and sympathy in HRI in care scenarios for older people, calling for an ontological turn in socio-gerontechnology “to embrace and handle ontological complexity” for robot use [27 p. 53].

However, making a social robot that effectively engages empathetically with users is a fraught ethical field, with opposition to, or at least ambivalence about, the value of a more emotionally engaging social robot. Studies of older people’s attitudes toward SARs show a high degree of variation in receptivity to companionship functions, with some studies reporting some older people seeing potential in SARs to alleviate loneliness [28][29][30][31][32]. Some also see danger in deception [30][32], or, as one researcher puts it, concerns that use of SARs “could lead to a dehumanized society” [33, p. 160] or other serious ethical issues [33][34][35][30][36][32][37].

These divergent views reflect a high degree of *cultural* ambivalence toward the prospect of using a robot for companionship, at least as they are currently presented, and indicate a strong aversion to techniques like enhanced empathetic verisimilitude that would potentially draw users into stronger emotional attachments with robots. A critical literature review conducted by [38] for instance, found that the second most common ethical issue discussed in the social robotics literature was concerns over deception, often deeming social robots to be inherently deceptive and hence morally suspect, given that their display of affect is synthetic and necessarily inauthentic – a view that was not dependent on any other negative consequences to an individual user. Users forming emotional attachments with social robots is often deemed to be laden with ethical risks, insofar as emotional dependency on a robot is deemed to undermine a user's autonomy and make them vulnerable to myriad harms, especially if users are already vulnerable [39].

Nonetheless, robots are increasingly designed to be anthropomorphized machines with emotive abilities, communicating with humans in collaborative social spheres [40][41][42][43][44].

### *D Alignment for empathetic robots*

Social science and humanities communication studies fields such as AI ethics, robot ethics, and philosophy delineate frameworks of value systems for human-robot interaction [45][46][47][48][49][50]. Media studies scholars contextualize the effects of deploying robots in societies by delving into platforms used to orchestrate digital communication, while also reflecting critically on the commercial hype that has for decades fueled a social robot industry [51][52]. People engage these discourses when establishing values about robots.

As human-AI relationships become more complex, what is further needed to ensure user safety is socioaffective alignment [13], such that "human goals and preferences become increasingly co-constructed through interaction with AI systems, rather than arising separately from them," which follows that "AI safety requires paying as much attention to the psychology of human-AI relationships as the wider societal factors and technical methods of alignment" [13, p. 4]. Considering this socioaffective context to achieve safe and aligned AI systems, they argue, will require: "a science of AI safety that studies real (not simulated) human-AI interactions in natural contexts and treats the psychological and behavioural responses of users as key objects of inquiry" [13, p. 14-15].

Popular culture impacts HRI adoption. Robots are adopted in tandem with cultural experiences that help establish social values. Film culture that depicts successful, empathetic robots in human-robot interaction scenarios provide an alluring incentive to achieve them in the real world. Films such as the Oscar nominated *The Wild Robot* [53] depict a future where robots learn to care for and champion living beings, exemplifying their emotional abilities rather than or at least equal to their rational prowess.

One research group explores the way film and animation techniques are helping to define how robots behave or emote in real world scenarios [54]. Along these lines, cinematic robots socialize the public to expect that real robots will genuinely emote in similar ways. A fictional humanoid robot such as Neill Blomkamp's Chappie of the film [55] by the same name is designed to emote empathy with humans to explore difficult social and political themes such as racism and state-sanctioned police violence. Chappie's characterization as a thinking, feeling, vulnerable, and kind robot amid a dystopian militarized world deliberately intends to inspire empathy, with the intent to get an audience to feel what Chappie feels.

In sum, social robots are incorporating AI systems that require humanities methods to achieve ethical alignment.

## IV Methodology and Methods

We have adopted a mixed method approach that combines two methodologies: *digital humanities archiving* using a collection of digital artifacts, and *rhetoric*, a theoretical approach that studies 'acts of persuasion'. As combined, the approach can then serve to improve designs for achieving the appearance of empathy. These combined methodologies constitute the first article contribution, a novel humanities-informed methodology to help achieve empathetic-appearing robots.

These methods were required due to previous findings by this team identified during six focus groups held in-person in Canadian community centers serving older people, and an exploratory online survey of 153 older participants fully documented in previous articles [1] [14]. A gap was identified: SARS claiming to be empathetic did not include design traits associated with empathetic care. A humanities approach to defining a human value system, empathy, was required.

### *A Digital humanities method for ethical alignment*

This section discusses a method to analyze cultural values *about* robots from a collection that indexes discourses informing those values. A digital humanities archive is sometimes compared to a digital museum. The purpose is to help people understand the cultural purpose of an artifact on display. Through their analysis, the researchers draw on this collection to determine which robot activities can be called empathetic. From this analysis, the authors contribute a novel and relevant *empathetic care vocabulary* that could be used by other designers/developers to help define robot empathy.

### *B Digital humanities collection: Humanoid Robots*

The researchers used an established digital collection called Humanoid Robots [56] to conduct this new study. The collection of 360 artifacts was interpreted according to the Dublin Core™ Metadata Initiative schema. The collection is housed in an established cultural analytics database called Fabric of Digital Life and it continues to grow.

Humanoid robots can include mobile robots, tabletop robots, and all categories of social robots that exhibit at least one humanlike physical embodied component. To conduct interpretation, the research archivists *identified* visual representations of robots found in inventor's papers and prototypes, science and technology research, journalism, video clips of fictional representations, and marketing materials for robot products (see figure 1).

*Parameters of the Humanoid Robots collection*

- collection established in 2017 and continues to grow
- artifact creation dates range in years 1927-2026
- 360 public artifacts at the time of writing
- 24 of 360 artifacts represent socially assistive robots (SARs)
- Each artifact assigned 16 metadata categories according to media type to define discourses, values, and other keyword categories.

*Scope of the Humanoid Robots collection*

To signify the scope of the Humanoid Robots collection, Table 1 provides a breakdown of the media types, including the largest video categories:

- 131 *video* artifacts are corporate videos from robot development companies
- 47 video artifacts are fictional film clips of humanoid robots
- 44 video artifacts are clips from news broadcasts that both inform on and sensationalize robots

The point is not to isolate these representations; it is to interpret them in discursive relations the way people (readers, viewers, etc.) experience them (e.g., in news items, advertising, films). People learn about robots and judge them through myriad discursive avenues.

Table 1 Media types in Humanoid Robots collection [56], source: Fabric of Digital Life

| Media Type: Video | Media Type: Text | Media Type: Graphic image | Media Type: Audio |
|---|---|---|---|
| Art (6)<br>Corporate Video (131)<br>Feature Film (47)<br>News Broadcast (44)<br>Personal Blog (3)<br>Television Show (16)<br>Video Game (2)<br>Video Lecture (8)<br>Web Series (10)<br>YouTube Clip (1) | Corporate Paper (6)<br>Government Publication (2)<br>Journal Article (6)<br>Magazine Article (45)<br>Newspaper Article (24)<br>Press Release (3) | Advertising (4) | Musical Song (1)<br>Radio Show (1) |

*Interpretive method*

Humanities researchers, professional archivists, and graduate students in Communication Studies tagged artifacts using the meta data scheme to classify and archive them in the public database according to several categories:

1. Publication/Creation Date
2. Creators/Contributors
3. Media Type (e.g., corporate video)
4. Persuasive Intent (e.g., advertising, information, fiction, etc.)
5. Discursive Type (e.g., invention, response to an invention, etc.)
6. Description (e.g., what appears in the artifact?)
7. HCI Platform (e.g., robotical, wearable, ambient, implantable, etc.)
8. Related Body Part (e.g., eye, ear, hand, etc.)
9. Relation To Body (e.g, around, on)
10. Augments (e.g., caring, protecting)
11. Technology Keywords (e.g., machine learning, facial recognition, etc.)
12. Marketing Keywords
13. General Keywords (e.g., Healthcare)
14. Source (e.g., website address, DOI, etc.)
15. Related Items (e.g., other artifacts by the same creators, cross-referenced items, etc.)
16. Related Collections

The database records automated metadata such as Date Archived, when it was Last Edited, and a citation line.

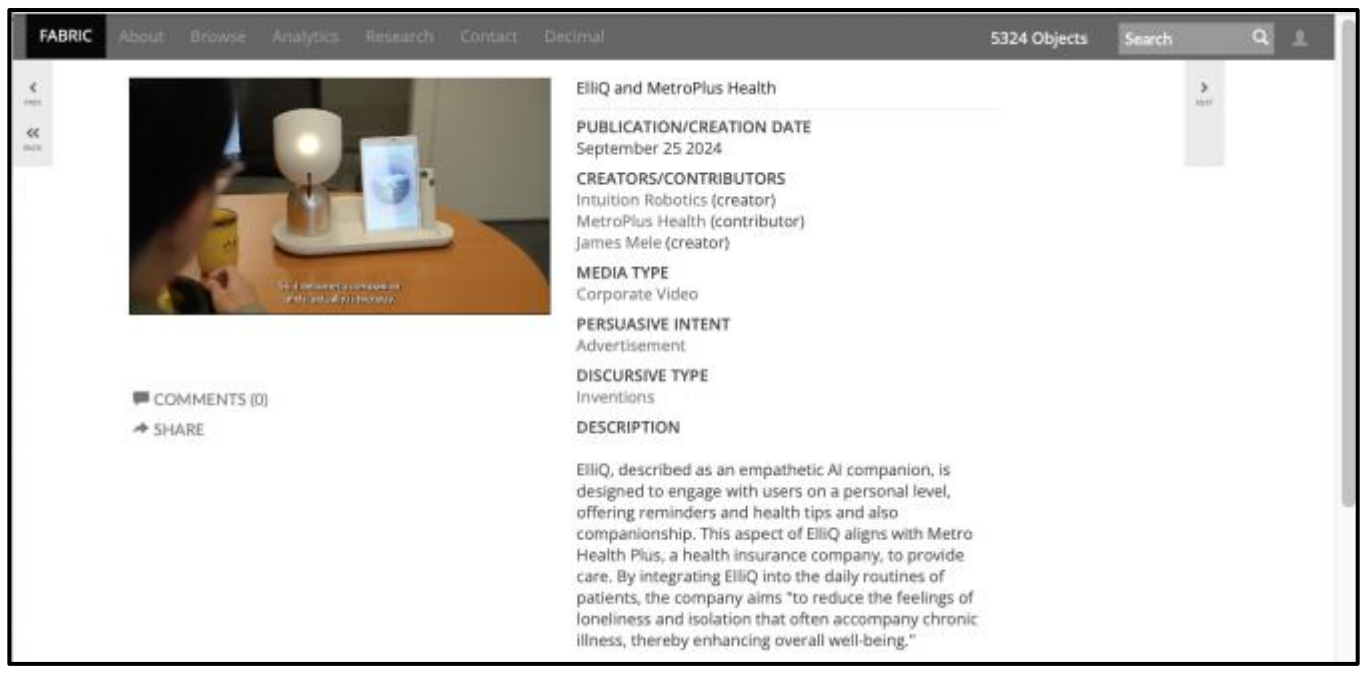

*Figure 1screenshot of an artifact on ElliQ that is described as an emotional companion robot from the Fabric of Digital Life database (photo permission, Isabel Pedersen)*

Choosing keywords for socio-technical systems and classifying them according to socio-ethical values requires interpretation. In keeping with a humanities interpretive approach, archivists respond to what they see in a graphic, video, or read in text. They designate interpretive, critical analysis of the cultural material in discursive relationships (figure 2). While the metadata scheme is highly standardized, the analysis deliberately promotes humanities interpretation for choices rather than a social science coding approach. For example, a researcher might interpret how a robot behaves in a 2014 news broadcast, interpreting empathy according to the technology at the time and values depicted in the video. Or they might question how a company depicts advertising robot empathy. Is the value of empathy used to *sell* a robot? These discursive markers influence the concept of robot empathy.

This method is a collaborative process. Subsequent archivists can return to artifacts to edit them, leading to a collective approach to classifying and then interpreting cultural artifacts in discourses. The editorial process does not override previous metadata choices (unless there is an error), and so the resulting collection signifies an ongoing collaboration and a source for cultural analysis of values that are always changing. All archivists' work and every metadata choice are recorded in the database, forming a history of contributions over the years leading to a collective, collaborative assessment of the artifacts.

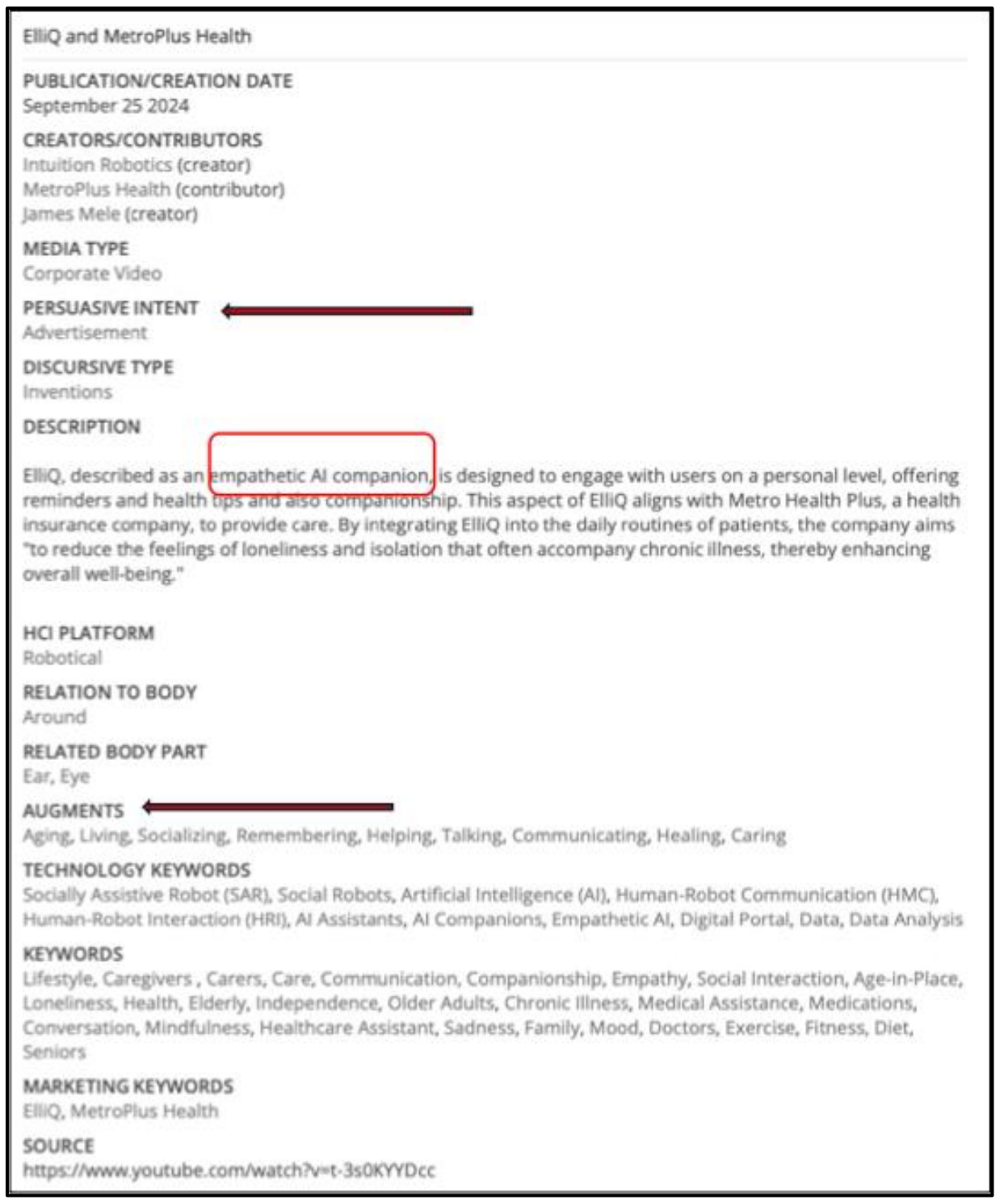

ElliQ and MetroPlus Health

PUBLICATION/CREATION DATE
September 25 2024

CREATORS/CONTRIBUTORS
Intuition Robotics (creator)
MetroPlus Health (contributor)
James Mele (creator)

MEDIA TYPE
Corporate Video

PERSUASIVE INTENT
Advertisement

DISCURSIVE TYPE
Inventions

DESCRIPTION

ElliQ, described as an empathetic AI companion, is designed to engage with users on a personal level, offering reminders and health tips and also companionship. This aspect of ElliQ aligns with Metro Health Plus, a health insurance company, to provide care. By integrating ElliQ into the daily routines of patients, the company aims "to reduce the feelings of loneliness and isolation that often accompany chronic illness, thereby enhancing overall well-being."

HCI PLATFORM
Robotical

RELATION TO BODY
Around

RELATED BODY PART
Ear, Eye

AUGMENTS
Aging, Living, Socializing, Remembering, Helping, Talking, Communicating, Healing, Caring

TECHNOLOGY KEYWORDS
Socially Assistive Robot (SAR), Social Robots, Artificial Intelligence (AI), Human-Robot Communication (HMC), Human-Robot Interaction (HRI), AI Assistants, AI Companions, Empathetic AI, Digital Portal, Data, Data Analysis

KEYWORDS
Lifestyle, Caregivers , Carers, Care, Communication, Companionship, Empathy, Social Interaction, Age-in-Place, Loneliness, Health, Elderly, Independence, Older Adults, Chronic Illness, Medical Assistance, Medications, Conversation, Mindfulness, Healthcare Assistant, Sadness, Family, Mood, Doctors, Exercise, Fitness, Diet, Seniors

MARKETING KEYWORDS
ElliQ, MetroPlus Health

SOURCE
https://www.youtube.com/watch?v=t-3s0KYYDcc

*Figure 2: A screenshot of the metadata for an artifact called ElliQ and MetroPlus Health that is described as an emotional companion robot from the Fabric of Digital Life database (photo permission, Isabel Pedersen)*

*C Empathetic care vocabulary*

One issue is that inventors call robots 'empathetic' without defining empathetic HRI behaviours. They are simply assumed.

The database provides an *augmenting* category to focus on what technologies can *do for humans*. This keyword field helps researchers classify human capabilities that a robot company or creator 'claims' a robot can perform or appear to perform to augment human abilities.

To focus on socially assistive robots, this team chose the keyword *caring* to help *define* empathetic robot activities in discourses. For sake of clarity, not all humanoid robots are designed for care, some might be designed for retail, manufacturing, or military usage and the *augmenting* keywords in these cases are quite different.

From the Humanoid Robots collection,

- 70 artifacts of the 360 total are tagged with the *Augments* keyword *caring*
- From the caring subset of 70, 99 additional augments keywords are co-present with *caring*

With this classification, the authors sought to identify a set of care activities cross referenced with other types of augmentations. The authors analyzed the keywords, reviewed the artifacts in discourses, and identified three care categories (see figure 3): physicalist, interactive, and empathetic. These categories are not mutually exclusive, as one humanoid robot artifact can exhibit all three e.g., by walking *(physicalist),* talking *(interactive),* and comforting *(empathetic)*.

The article here contributes a design vocabulary, a set of words to define whether a robot is *being empathetic* to humans, thus creating an *empathetic care* vocabulary.

To this end, the authors pull 15 specific *augments* empathetic keywords (see figure 3). They are: bonding, caring, comforting, expressing, fearing, feeling, grieving, healing, imagining, knowing, loving, perceiving, remembering, sharing, and understanding.

| | | | |
|---|---|---|---|
| Accessing (1) | Engaging (5) | Lifting (1) | Seeing (2) |
| Adapting (2) | Entertaining (1) | Listening (5) | Sensing (1) |
| Aging (36) | Evolving (1) | Living (24) | Serving (1) |
| Assisting (22) | Exercising (1) | Loving (4) | Sharing (4) |
| Attending (1) | Exploring (1) | Making (1) | Shopping (3) |
| Bonding (11) | Expressing (4) | Managing (5) | Singing (1) |
| Caring (67) | Fearing (2) | Monitoring (1) | Sleeping (1) |
| Carrying (2) | Feeling (16) | Moving (11) | Socializing (19) |
| Catching (1) | Fighting (2) | Navigating (3) | Solving (1) |
| Choosing (1) | Gesturing (1) | Organizing (2) | Speaking (10) |
| Cleaning (3) | Grabbing (1) | Perceiving (2) | Supporting (3) |
| Climbing (1) | Greeting (2) | Performing (1) | Surveilling (3) |
| Collaborating (1) | Grieving (2) | Playing (5) | Talking (12) |
| Comforting (4) | Guiding (1) | Preparing (1) | Teaching (6) |
| Communicating (39) | Healing (3) | Preventing (1) | Throwing (1) |
| Connecting (23) | Helping (22) | Projecting (1) | Tracking (1) |
| Controlling (2) | Imagining (1) | Protecting (9) | Training (1) |
| Cooking (5) | Imitating (2) | Purchasing (1) | Treating (5) |
| Creating (1) | Improving (7) | Recovering (1) | Understanding (6) |
| Dancing (1) | Innovating (1) | Relating (2) | Viewing (1) |
| Deciding (1) | Instructing (1) | Relieving (1) | Walking (4) |
| Designing (1) | Interacting (29) | Remembering (10) | Washing (1) |
| Dressing (1) | Killing (3) | Reminding (3) | Watching (1) |
| Eating (1) | Knowing (1) | Responding (3) | Working (9) |
| Emoting (1) | Learning (12) | Searching (1) | |

⇒ Physicalist care vocabulary, e.g., working, shopping, cooking, walking
⇒ Interactive care vocabulary, e.g., speaking, collaborating, listening
⇒ Empathetic care vocabulary, e.g., knowing, understanding, remembering

*Figure 3 a list of the frequency of Augments keywords in the Humanoid Robots Collection, colour-coded according to care vocabulary types*

## V Rhetorical Empathy for Human-robot Interaction Design

While the last section defines robot activities that *are empathetic*, this section defines how robots *might appear empathetic* based on human empathy.

This section uses Rhetoric as a theoretical framework, an interpretative lens. Rhetoric is a method in communication studies and specifically semiotics that deals with 'acts of persuasion'; it explores how minds are changed or actions taken to influence people or organizations through media of communication (e.g., public speeches, propaganda, news, chatbots, film, Instagram pages, misinformation, etc.).

In this part, we concentrate on definitions of empathy and aspects of empathy for the purposes of HRI design.

### *A Empathy defined*

There are many definitions for the word, empathy. From the Greek root empatheia (em- 'put into' + pathos 'pity or compassion'), empathy is the act of understanding, being aware of, being sensitive to, and vicariously knowing the feelings, thoughts, or experiences of another. Often described as a key feature of meaningful social interaction, "empathy is a process in which an observer vicariously shares the emotion or intention of another person and thereby understands what this other person feels or intends" [57, p. 41]. Its simplicity is deceptive, and empathy scholars are quick to point out that the process, mechanism, or journey to gain empathy is no simple experience [57].

Gaining empathy then functions as a rhetorical strategy in different contexts [58] [59].

### *B Function of empathy*

When one discusses *robot empathy*, the term requires assessment according to how it meaningfully functions. Empathy then is both a social value (e.g., it is good to empathize with people) and it has a rhetorical, potentially manipulative function (e.g., persuading people to adopt a political view through friendship), and so also has a design dimension (e.g., embedding persuasive components in computer interfaces).

Based on the methods used and the theoretical focus in this section, we argue for three design concepts to meet the objective of techniques aligned with the social value of empathy, and avoiding performed empathy's manipulative potential. It serves as an interdependent design criterion for consubstantial human-robot interaction (see Figure 4). Discussed in the section below, the first, consubstantiality, takes a primary role for the other two, personification and dialogism.

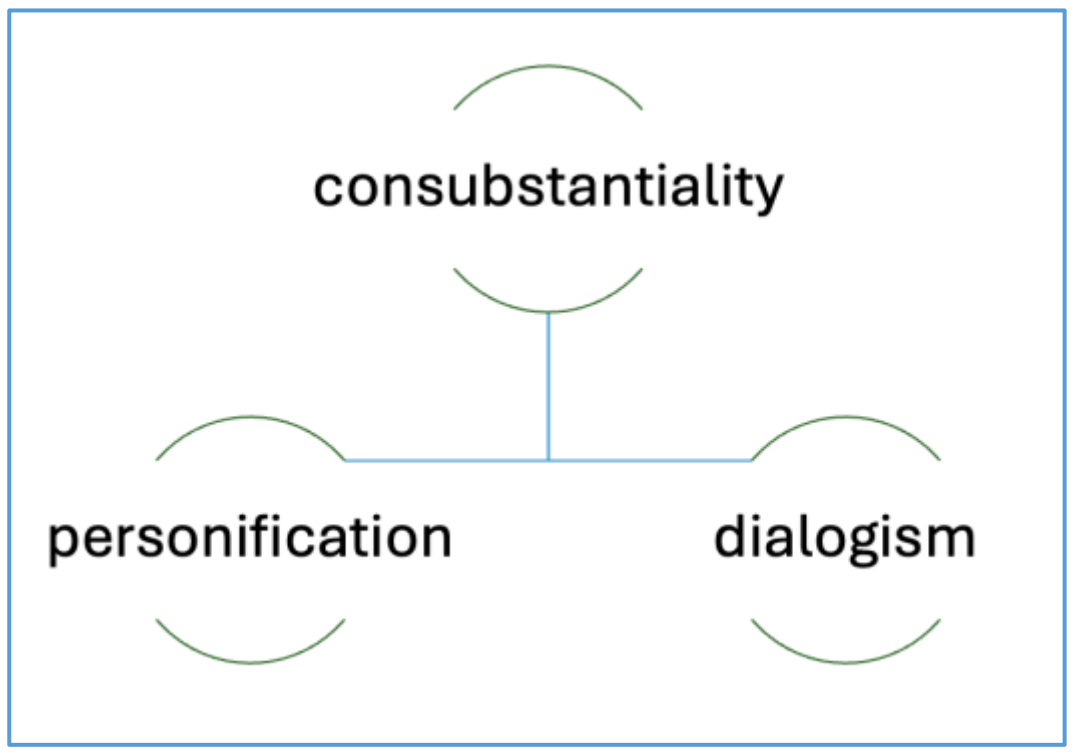

*Figure 4 A tree structure line drawing of an interdependent design criteria for human-robot interaction, with Consubstantiality at the head, Dialogism and Personification serving as branches.*

*C Consubstantiality – designing personal robots that appear to have mutual respect in robot interactions*

Empathy as a social value requires mutual respect. Humans are naturally divided from each other. Any undertaking to cooperate, collaborate, act through interdependence and establish mutual respect are constituted by what we call *consubstantiality*. We break down consubstantiality into three further design concepts: identification, context, and dignity.

*D Empathy requires identification*

Identification – that is, seeing oneself in another – is the means through which one becomes consubstantial with one another [60]. The promise of empathy in close human relationships is that it offers the condition of being known by another, as when someone simply 'gets you.' The anticipated social bond involves holding selves and others in separate, but consubstantial relationships. Therefore, empathy is a desired quality for any relationship because the labour of telling, explaining, or defending aspects of the self, accrues toward an ideal end when one is *known*. Empathy, then is functional to the design of satisfying social interaction and socioaffective alignment.

*E Empathy requires context*

Without context, social interaction cannot occur in meaningful ways in human-computer interaction and human-robot interaction. Put another way, expressing certain words without knowledge of scope and proper context of the other person's life, does not produce the grounds for respect needed for a beneficial relationship. Context is more than the actual words in a particular conversation. One scholar writes, "we place the object of our concern within a setting of particular scope. [A context that]… indicates where the explanation stops; where it satisfies the terminological cluster available" [61, p. 16].

In one of the previous survey responses to the use of social robots in homes, one older person expresses: "with such a vast amount of people feeling 'trapped' or depressed [a social robot] would help seniors ward off loneliness and social isolation, creating better brain function/activity and allow them to learn new ways of doing things." Achieving an ideal scenario such as this would require context through a deep ontological understanding of how a person feels, learns, and appreciates activities, all contextualized within that person's life, unlike current robot designs that offer a catchall of social features from an often-stereotypical view of older people.

*F Empathy recognizes the core value of dignity*

If one accepts that SARs can ethically perform the functional appearance of social bonding, 'acting empathetically', then they must be designed to appear to feel what the other feels, and not solely base responses on the raw content of conversations.

Of course, empathy implies its antonym, apathy that comes from the Greek apathēs without feeling or emotion (pathos). The threat of an apathetic partner in a relationship is menacing because one will have to experience a social disconnect or be alone. Indeed, inadequate or unconvincing performance of consubstantiality opens a robot to the ethical hazard of objectification, turning the user into an object to be acted upon, rather than an interlocutor. Consubstantiality, then can be used to judge robot behaviour that appears alienating or apathetic.

*Consubstantial approaches to remembering*

Helping older people remember aspects of their lives is a prominent theme in gerontology discourse. The authors viewed all the metadata that informs the claim that a humanoid robot has the technical capability to perform memory augmentation. This includes technology categories relating to *Remembering*, including face tracking, speech recognition, and object detection.

Yet, few of these artifacts aligned with helping a user remember with empathetic consubstantiating gerunds like imagining, perceiving, and expressing. A consubstantial approach would involve helping a user remember through *identification.* The team's analysis determines that current social robot designs are lacking the means to demonstrate this key aspect of empathy.

Traits of empathetic consubstantial robot interaction for *remembering should include:*

- Identifying with a person's life experiences, ensuring that the AI has enough context to perform respectful listening
- Displaying features that 'behave' as if a social bond has been established through highly contextualized personal representations signifying human memory

- Helping an individual imagine, explore, regret or delight in one's memories in a purely personal way

How can consubstantiality be incorporated in design? From consubstantiality's three criteria – identification, context, and the appearance of respectful social bonding – the design concepts Dialogism and Personification provide a better design foundation.

*G Dialogism – designing the appearance of dynamic humanlike dialogue in human contexts*

Dialogic processes refer to literal and implied meanings spoken or visualized by communicators and interpreted by listeners/viewers in relational, contextualized, discursive exchanges. With the advent of LLMs, social robots are being designed to engage in lively realistic conversations with people. Part of the importance of AI systems is the idea that empathetic robot responses will become personal and dynamic in conversational exchanges creating the feeling of interpersonal socioaffectivity [13]. Going forward, the deployment of AI chatbots as conversational agents will further inform the technologies used to make social robots and SARs more interactive and personable.

*Dialogism: Speaking and Listening*

While the ability to engage in dialogue with a user is a common feature of SARs, those in the *Humanoid Robot* collection reveal a limited number of empathetic activities associated with conversation in social robots advertising. In one example, the ElliQ tabletop robot by Intuition Robotics (see figures 1 and 2), is sold on the premise that the robot provides older people with a companion in the home that proactively initiates conversation.

There are 16 artifacts that relate to ElliQ in the Humanoid Robots collection that have been tagged with *augments* keywords. Most of ElliQ's extensive conversational abilities expressed in the artifacts do not significantly align with *empathy*. Of these, only *caring*, *healing*, *feeling*, and *remembering* fall in the empathetic care vocabulary list They largely fall in the interactive care vocabulary category demonstrating social functions rather than empathetic ones.

Traits of empathetic dialogic robot interactions include:

- A more productive dialogic trait might be to design a robot to refrain from speaking where empathy calls for it. That is, *not* speaking at times when listening, to demonstrate identification with a person.
- Using language to serve an interpersonal contextualized function, for example, when empathy requires helping a person *imagine*, *express themselves*, or help them come to terms with and *understand* loss through listening.

In the previous section, we discuss how empathy offers the condition of being known by another, when someone simply 'gets you'. This design pathway suggestion is not intended to critique ElliQ, but rather to illustrate an idealized design concept for empathetic HRI appropriate to a companion role.

*Dialogism: Seeing and Being Seen*

Researchers and inventors work to embed the appearance of emotional signs in functional AI applications through embodied meaning-making and interpersonal communication. For example, many commercial robots (or those in commercial development) have facial recognition capabilities so they can identify and specifically address the people they interact with by recognizing individuals through stored biometric data.

Likewise, robots also have pre-built facial expressions in order to respond in a socially expected manner during these exchanges, including having robots smile or frown. More advanced expressive repertoires are under development, including those related to a robot's gaze (e.g. the ability to avert a gaze and so hold or look away from the user's gaze) is undergoing development [44].

Traits of empathetic dialogic robot interactions include:

- Dynamic dialogue effected by incorporating corresponding physical cues, expressions and gaze, as a means toward ethical consubstantial human-robot interaction, provided that it is linked with empathetic design features.

*H Personification – designing personal robots that appear to have human-like personalities*

Personification is the final design concept that requires ethical and cultural alignment for empathy in HRI. Part of this design conceptualization, however, requires ameliorating *deceptive* design. *To personify* is to represent a thing in the form of a person. Personification is a figure of speech that is used rhetorically to persuade/deceive an audience to believe a non-human thing has humanlike characteristics. Closely related, anthropomorphism is a cultural phenomenon involving the attribution of human-likeness to non-human entities. One team of researchers explain that "Anthropomorphic perceptions usually arise unconsciously when a non-human entity bears enough resemblance to humanness to evoke familiarity, leading people to interact with it, conceive of it and relate to it in ways similar to as they do with other humans" [12 p. 93]. Humanoid robot design exploits the human tendency toward anthropomorphism, and so its ethical positioning must be carefully managed.

Yet, AI as a field is systemically described as anthropomorphic; "classically, following the work of Alan Turing, human-likeness was the operative standard in definitions of AI. A system could only be held to be intelligent if it could think or act like a human with respect to one or more tasks" [62 p. 629]. Whether overt or reticent, AI is usually legitimized by an economic imperative, which is for AI to mimic (and ultimately better) human behaviour in order for machines to optimize or to gain some apparent advantage over humans in performing tasks (e.g., cheaper, faster, smarter, etc.). HRI design then needs to pinpoint personified activities and educate users when they are deployed to avoid deceptive outcomes.

*Personification: Emoting and remembering emotions*

While robots are generally still clunky and awkward in both their gestures and expressions, there is research momentum to achieve responsive physical and expressive interactions [44]. As their artificial bodies become more ambulatory, flexible, and even soft, robots will be increasingly capable of appearing empathetic.

In the analysis of the collection, marketing videos for SARs often provide at least hints of human-like personalities, making them appear less like pieces of machinery and more like robot characters performing a human-like role in a user's life. ElliQ, mentioned above, is designed to display basic head and neck gestures to imply curiosity and attentiveness, and is depicted with a clear, calm female voice. Engineering Arts's Ameca, a standing robot with deliberate character development, similarly is portrayed exhibiting facial expressions that read as curious and attentive. Still in prototype, Ameca combines verbal, facial, and personified expressions generated using the GPT-3 LLM [63].

More sophisticated forms of personification currently available via AI chatbots are making their way into social robots, for instance via Furhat's tabletop robot, that uses a back-projected 3D face screen to customize the robot's face to a user's specifications, and the capacity to incorporate a wide selection of voices (including those that clone the voices of real people), integrated with Microsoft Azure and Google Cloud speech recognition models.

Drawing on the empathetic care vocabulary and activities, personified robots will begin to perform more personal, cultural roles like assisting with *grieving* in HRI. The growing industry of "grieftech", which at present mainly creates chatbots that mimic a person who has died and allows their surviving loved ones to continue talking "with" them, will also surely accelerate the personification capacities of SARs. Many observers are highly skeptical that such bots can be ethically created and used [64], despite their primary focus on enacting empathy [65][66]. A starting point is mediating highly personal contexts for users with consubstantial design goals.

Traits of empathetic personified robot interactions include:

- Provide accurate responses to physical and social surroundings, which will redound with a robot's awareness of a particular user's emotional needs with context.
- Avoid stylized robot characters drawn from stereotypes and pre-conceived profiles of older people.
- Customize a robot personality according to a user's desired level of empathetic activities. Provide highly personal attention to bonding, comforting, expressing, fearing, feeling, grieving, healing, imagining, knowing, loving, perceiving, remembering, sharing, and understanding, with attention to cultural expectations as a starting point.

## VI Conclusion

Trustworthy, ethical robot interaction with older people needs to establish proper design goals rather than generic acquiescence to the status quo. There is a great commercial impetus for robots and other AI agents to progress to perform empathetic acts for older adults in mainstream cultural spheres, and each year robot industries are bringing us closer toward that goal.

This article has contributed a (1) novel humanities-informed methodology for designing empathetic HRI, (2) an empathetic care vocabulary that could be used by other designers/developers to characterize empathetic activities, and (3) value-based design concepts and design traits for developers to help achieve socially assistive AI-based robots for older people that behave with empathy.

More design-focused work needs to originate from humanities fields. The Alan Turing Institute, Britain's national institute for data science and artificial intelligence, has noted a foundational gap in the lack of humanities methods for AI development, "AI systems increasingly produce and act upon cultural outputs – language, images, narratives – yet lack frameworks for interpreting the cultural content they generate and encounter" [67 p. 5]. Further it notes: "These interpretive challenges fall precisely within the expertise of the humanities, arts, and qualitative social sciences – disciplines that specialise in understanding cultural meaning, contextual nuance, and interpretive complexity" [67 p. 5].

Future research on design traits for AI robots and empathy could engage more deeply with extensive foundational theoretical work from posthuman studies [68], history and aesthetics of cybernetics [69], and other technocultural humanities fields.

Industry goals should involve satisfying older users, who benefit from emotionally supportive functions, but also to help them avoid the pitfalls of potential manipulation that occurs when careful design values are ignored.

## Acknowledgment

We thank our graduate student research assistants.